\documentclass{article}

\usepackage{newtxtext,newtxmath}

\usepackage[T1]{fontenc}
\usepackage{ae,aecompl}

\usepackage{graphicx}	
\usepackage{amsmath}	
\usepackage{amssymb}	
\usepackage{natbib}
\usepackage{graphicx}
\usepackage{journals}
\usepackage{lscape}
\usepackage{xcolor}
\usepackage{mathtools}
\usepackage[cal=boondoxo]{mathalfa}
\usepackage{authblk}

\newcommand{\figref}[1]{Fig.~\ref{#1}}
\newcommand{\refp}[1]{(\ref{#1})}
\newcommand{\Figref}[1]{Fig.~\ref{#1}}
\newcommand{\eqnref}[1]{equation~(\ref{#1})}

\newcommand{\tabref}[1]{Table~\ref{#1}}
\newcommand{\Tabref}[1]{Table~\ref{#1}}
\newcommand{\secref}[1]{Sect.~\ref{#1}}
\newcommand{\Secref}[1]{Sect.~\ref{#1}}
\newcommand{\appref}[1]{App.~\ref{#1}}

\newcommand{\be}{\begin{equation}}
\newcommand{\ee}{\end{equation}}
\newcommand{\eep}{\;\;.\end{equation}}
\newcommand{\eec}{\;\;,\end{equation}}
\newcommand{\bea}{\begin{eqnarray}}
\newcommand{\bel}[1]{\be\label{#1}}

\newcommand{\eea}{\end{eqnarray}}

\newcommand{\tp}[1]{\mbox{$\times10^{#1}$}}
\newcommand{\nablav}{{\mathbf \nabla}}
\newcommand{\curl}{\nablav\times}
\newcommand{\rv}{\mathbf r}

\newcommand{\uvp}{\hat{\mathbf \phi}}
\newcommand{\uvz}{\hat{\mathbf z}}

\newcommand{\Uv}{\mathbf{u}}
\newcommand{\Bv}{\mathbf{B}}
\newcommand{\jv}{\mathbf{j}}

\newcommand{\ST}{\mathcal{S}}

\newcommand{\cj}{j^\star}
\newcommand{\cjln}{\cj_{\ell n}}

\newcommand{\kln}{k_{\ell n}}
\newcommand{\Pl}{P_{\ell}}
\newcommand{\Plt}{\Pl(\theta)}
\newcommand{\DSG}{DSG}
\newcommand{\other}[1]{\tilde{#1}}
\newcommand{\back}[1]{\overline{#1}}
\newcommand{\dist}[1]{{#1}^\prime}
\newcommand{\gravc}{G}
\newcommand{\GF}{\Gamma}
\newcommand{\pot}{\Psi}
\newcommand{\potb}{\back{\pot}}
\newcommand{\potp}{\dist{\pot}}
\newcommand{\rhob}{\back{\rho}}
\newcommand{\rhop}{\dist{\rho}}
\newcommand{\rhoU}{\rho^U}

\newcommand{\pb}{\back{p}}

\newcommand{\pp}{\dist{p}}

\newcommand{\gb}{\back{g}}

\newcommand{\gv}{\mathbf{g}}
\newcommand{\gvb}{\back{\mathbf{g}}}

\renewcommand{\Gamma}{\varGamma}
\renewcommand{\Psi}{\varPsi}

\title{Linking Zonal Winds and Gravity: 
The Relative Importance of
Dynamic Self Gravity}

\author[1]{J. Wicht}
\author[1]{W. Dietrich}
\author[1]{P. Wulff} 
\author[1]{U.~R.~Christensen} 

\affil[1]{Max Planck Institute for Solar System Research, 
Justus-von-Liebig-Weg 3, 37077 G\"ottingen, Germany}

\begin{document}

\maketitle

\begin{abstract}
Recent precise measurements at Jupiter's and Saturn's gravity fields 
constrain the properties of the zonal flows in the outer envelopes 
of these planets. 
A simplified dynamic equation, sometimes called 
the thermal wind or thermo-gravitational wind equation, 
establishes a link between zonal flows and the 
related buoyancy perturbation, which in turn can be exploited to
yield the dynamic gravity perturbation. 
Whether or not the action of the dynamic gravity perturbation   
needs to be explicitly included in this equation,  
an effect we call the Dynamic Self Gravity (\DSG),  
has been a matter of intense debate. 
We show that, under reasonable assumptions, 
the equation can be solved (semi) analytically. 
This allows us to quantify the impact of
the \DSG\ on each gravity harmonic, 
practically independent of the zonal flow or the details 
of the planetary interior model. 
The impact decreases with growing spherical harmonic degree $\ell$. 
For degrees $\ell=2$ to about $\ell=4$, the \DSG\ is a 
first order effect and should be taken into account in any
attempt of inverting gravity measurements for zonal
flow properties. For degrees of about $\ell=5$ to roughly $\ell=10$,  
the relative impact of \DSG\ is about $10$\% and thus seems 
worthwhile to include, in particular since this comes at little 
extra costs with the method presented here. 
For yet higher degrees, is seems questionable whether 
gravity measurements or interior models will ever reach 
the required precision equivalent of the \DSG\ impact of 
only a few percent of less. 
\end{abstract}




\section{Introduction}

For the first time, the high precision of gravity measurements 
by the Juno mission at Jupiter and the Cassini Extended Mission at 
Saturn allow the detection of the tiny perturbations related to the 
fierce zonal winds in the outer envelopes. 
However, there is an ongoing dispute about the appropriate equation 
for linking gravity perturbations and zonal flows 
\citep{Cao2017,Kong2018,Kaspi2018}. 
A particular matter of debate is whether the back-reaction of the 
gravity perturbations on the flow dynamics has to be taken into account.
This article addresses the question with a new semi-analytical approach. 

The impact of gravity on the flow dynamics is generally given by the 
Navier-Stokes equation. The hydrostatic solution decribes  
the zero order balance between pressure gradient 
and effective gravity that defines the fundamental 
background state. 
The effective gravity is the sum of gravity and 
the centrifugal force due to the planetary rotation. 
Respective equipotential surfaces coincide with 
surfaces of constant pressure and density and 
different methods have to devised for finding the 
respective solution 
\citep{Zharkov1978,Wisdom1996,Hubbard2013,Nettelmann2017}.


The centrifugal forces lead to a spheroidal deformation 
of equipotential surfaces and density distribution
$\rho$. 
The gravity potential
\bel{eq:VC}
  \pot(\rv) = -\frac{\gravc M}{r}\;
  \left[ 1 - \sum_{\ell=2}^{\inf}\,J_\ell\;\left(\frac{R}{r}\right)^\ell\;P_{\ell}(\theta)
  \right]
\ee
thus acquires equatorially symmetric contributions of even
degree $\ell=2 n$ with $n=1,2,3,...$. 
Here $\gravc$ is the gravity constant, $M$ the planetary mass,
$R$ the planetary radius, $\theta$ 
the colatitude, and $P_\ell$ a Schmitt-normalized Legendre 
Polynomial of degree $\ell$. 
The gravity harmonics $J_\ell$ are given by the volume integral 
\bel{eq:GM}
J_\ell = \frac{2\pi}{M R^\ell}\;\int\,d\,V\;{r}^{\ell}\;
\rho(r,\theta)\;P_{\ell}(\theta) 
\ee
and describe deviations from the spherically symmetric 
zero order contribution.

The degree of rotational deformation 
depends on the relative importance of centrifugal forces
to gravity, which can be quantified by $q=\Omega^2/(\gravc\rho)$, 
where $\Omega$ is the planetary rotation rate.
For Jupiter, $q$ remains below $0.1$ and 
deviations from the spherically symmetric gravity thus amount 
to only about $5$\%. 
For Saturn, $q$ is about two times larger than for 
Jupiter, which is consistent with the stronger deformation of the 
planet. Since gravity 
mostly originates from the higher densities in the deep interior,
where the deformation is smaller, the deviation of spherical 
gravity is only slightly larger than for Jupiter. 

Some of the classical methods for solving
the rotationally deformed hydrostatic solution can be extended to 
include geostrophic zonal flows, which depend only on the 
distance to the rotation axis 
\citep{Hubbard82,Kaspi2016,Wisdom2016,Galanti2017,Cao2017}. 
\citet{Cao2017} explore geostrophic zonal flows that are reminiscent 
of Jupiter's equatorial jet. 
They report that the zonal wind induced gravity  
amounts to only three permil of the gravity induced 
by the planetary rotation for $J_2$. For $J_8$, both effects
have a comparable magnitude, while zonal wind effects
dominate for larger degrees. For $J_{20}$, the 
related contribution is ten orders of magnitude larger than 
its rotational counterpart. 

\citet{Cao2017} point out the the small contributions at 
low degrees can easily
be offset by uncertainties in the background model, 
for example the composition, the equation of state, 
or the presence of stably stratified layers \citep{Debras2019}.
In practice, the even harmonics up to $J_4$, possibly 
even $J_6$, serve to constrain the zero
order background state. 
Only contributions beyond $J_6$ could thus 
reliably be exploited to gain information on the 
equatorially symmetric zonal flows.

The situation changes for the equatorially antisymmetric 
gravity harmonics, which can be interpreted directly 
in terms of a first order dynamic perturbation. 
(The hydrostatic background state being equatorially symmetric
and of zero order.)
The effect of non-geostrophic flows is estimated based 
on a simplified dynamic balance. Viscous forces are negligible 
in the Gas giant atmospheres. Since the zonal 
winds are rather stable and significantly slower than 
the planetary rotation, inertial forces are also  
significantly small than Coriolis forces, buoyancy, 
or pressure gradients.
When taking the curl of the force balance, 
the pressure gradient also drops out and 
the first order balance reads 
\bel{eq:FOB}
2\Omega\;\frac{\partial \rhob\,U_\phi}{\partial z} = 
\uvp\cdot\curl \;\left( \rhop\,\nablav\potb_e + \potp\,\nablav\rhob\;\right)
\eec
where $z$ is the 
distance to the equatorial plane, 
$\potb_e$ the effective background potential, $\rhob$ 
the background density, $\rhop$ the density perturbation and 
$\potp$ the gravity perturbation.
Note that we have also neglected the Lorentz-force related term
here. While Lorentz forces may play a significant role 
at depth where electrical conductivities are higher, 
the are much less important in the outer envelope 
where zonal flows are fast but electrical conductivities 
drop to zero. 
  
An important point of debate is whether the term involving 
the gravity perturbation $\potp$ yields a significant 
contribution or can be neglected. 
We refer to this term as the Dynamic Self Gravity (\DSG) here. 
When the \DSG\ can be neglected, the balance \refp{eq:FOB} reduces to the 
classical Thermal Wind Equation (TWE).
The full balance including \DSG\ has thus been called Thermo-Gravitational
Wind Equation (TGWE) by \citet{Zhang2015}. 

One group of authors insists that the \DSG\ term can be as large as 
the term involving $\rhop$ 
\citep{Zhang2015,Kong2016,Kong2017,Kong2018}. They also point out that 
neglecting the \DSG\  would fundamentally change the mathematical 
nature of the solution. 
To explore the \DSG\ impact, \cite{Kong2017} assume a zonal wind system
that reproduces the observed equatorially antisymmetric 
winds at Jupiter's cloud level and retains a geostrophic wind morphology at 
depth, i.e. the morphology is continued downwards 
along the direction of the rotation axis. 
Their amplitude, however, is supposed to decay linearly 
with the distance to the equatorial plane $z$. 
They report that neglecting the \DSG\ has  
a surprisingly large impact on $J_1$, and reduces $J_3$, 
$J_5$, and $J_7$ by $25$\%, $15$\%, and $7$\%, respectively.  

A second group of authors argues that the 
\DSG\ can be neglected 
\citep{Kaspi2016,Galanti2017,Kaspi2018,Iess2019}. 
\citet{Galanti2017} explore a simplified equatorially 
symmetric zonal flow system that matches the main features of 
the respective flows at cloud level. The wind structure is again 
continued downward along the rotation axis, but assuming an additional
exponential decay with depth.  
They conclude that the \DSG\ has only a minor
impact. However, their figure 6 suggests that the 
zonal-flow-related 
$J_2$ decreases by up to $100$\% when neglecting the \DSG.

\citet{Guillot2018} use Jupiter's even gravity harmonics 
up to $J_{10}$ measured by the Juno mission to constrain
the planets equatorially symmetric zonal winds. 
Analyzing a suit of possible background models, they
report that  $J_6$, $J_8$ and $J_10$ can only be explained 
when the perturbation related to the zonal winds is taken
into account. Using the TWE and assuming the exponentially 
decaying wind structure by \citet{Galanti2017}, \citet{Guillot2018} 
report that the e-folding depth lies somewhere 
between $2000\,$ and $3500\,$km. 

The odd gravity harmonics $J_3$ to $J_9$ based on Juno measurements 
were also recently used to constrain the depth 
of the zonal winds.
\citet{Kong2018} use the full TGWE equation while 
\citet{Kaspi2018} neglected the \DSG. 
Both articles where roughly able to explain the gravity 
harmonics with equatorially antisymmetric zonal winds 
that reproduce the observed surface winds. Both also conclude
that the winds must be significantly slower than observed at the 
surface below a depth of about $3000\,$km. 
However, the suggested radial profiles differ significantly. 
Since the results rely on different interior models, 
methods, and assumed zonal flow profiles, 
it is difficult to judge to which
to degree the results are influenced by the \DSG. 

\citet{Iess2019} explore Saturn's even gravity harmonics 
$J_2$ to $J_{10}$ measured by the Cassini mission. 
Like for Jupiter, $J_6$, $J_8$ and $J_{10}$ can 
only be explained when considering the zonal wind impact. 
However, unlike for Jupiter, 
a slight modification of the surface wind structure  
is required. \citet{Iess2019} report that these
modified winds reach down to a depth of about $9000\,$km.
While generally using they TWE approximation, \citet{Galanti2019} 
report that $J_8$ and $J_{10}$ increase by about $10$\%
when including \DSG\ in the TGWE approach.
\citet{Galanti2019} in addition also analyze the odd harmonics 
$J_3$ to $J_9$ and confirm the inferred 
depth of Saturn's zonal winds. 

Here we explore the relative importance of the \DSG\ 
with a new (semi) analytical method.  
\Secref{sec:equations} introduces the differential equations 
that define the gravity potential. \Secref{sec:method} then 
develops the solution method. \Secref{sec:solution}
discusses solvability aspects with some illustrative solutions and 
\secref{sec:impact} quantifies the relative impact of \DSG. 
The paper closes with a discussion in \secref{sec:discussion}. 

\section{From Navier-Stokes Equation to\\ Inhomogeneous Helmholtz Equation}
\label{sec:equations}

The link between the dynamics and gravity is provided by 
the Navier-Stokes equation
\bel{eq:NS}
\rho \left(\frac{\partial}{\partial t} + \Uv\cdot\nablav \right)\,\Uv 
+ 2 \varOmega \rho\;\uvz\times\Uv = -\nablav p + \rho\,\gv_e + \jv\times\Bv + 
\nu\,\nablav\cdot \ST
\eec
where $\Uv$ is velocity, $\uvz$ the unit vector in the direction 
of the rotation axis, $p$ the pressure, $\jv$ the electric current, 
$\Bv$ the magnetic field, $\nu$ the kinematic viscosity, 
and $\ST$ the traceless rate-of-strain tensor for 
constant kinematic viscosity:
\bel{eq:ST}
 \ST= \rho\left(\frac{\partial u_i}{\partial x_j} +
                \frac{\partial u_j}{\partial x_i} -
                \frac{2}{3}\delta_{ij}\nablav\cdot\Uv \right)
\eep

The effective gravity $\gv_e$ can be expressed by an  
effective gravity potential,
\bel{eq:EG}
 \gv_e= -\nablav \pot_e = - \nablav \left( \pot + \pot_\Omega \right)
\eec
which is the sum of the gravity potential obeying 
the Poisson equation
\bel{eq:poiss}
 \nabla^2 \pot = 4\pi \gravc\;\rho
\ee
and the centrifugal potential
\bel{eq:CP}
 \pot_\Omega = - \frac{1}{2}\;\Omega^2 s^2
\eec
with $s=r\sin{\theta}$ being the distance to the rotation axis. 

The zero order force balance is given by the hydrostatic 
equilibrium with vanishing flow and magnetic field:
\bel{eq:HS}
 \nablav \pb = -\rhob\;\nablav \potb_e 
\eec
\bel{eq:potHS}
 \nabla^2 \potb_e = \left( 4 \pi \gravc\;\rhob - \Omega^2 \right)
\eep
Overbars mark the hydrostatic and non-magnetic background state, 
while primes denote the perturbation,  
except for flow and magnetic field. 

Linearizing with respect to the perturbations yields 
\begin{multline}
\label{eq:NSP}
\rhob \left(\frac{\partial}{\partial t}  + \Uv\cdot\nablav \right)\,\Uv +
2 \varOmega \rhob\;\uvz\times\Uv = -\nablav \pp - 
\rhob\,\nablav \potp_e 
\rhop\,\nablav \potb  
+ \jv\times\Bv + \nu\,\nablav\cdot \ST\;\;,
\end{multline}
\bel{eq:potp}
 \nabla^2 \potp = 4\pi \gravc\;\rhop
\eep

The linearized buoyancy term has two contributions, one due to 
the density perturbation and a second one due to 
the perturbation in gravity.
The latter can be separated into a conservative part,   
written as a gradient, and the remaining contribution:
\bel{eq:GP}
\rhob \nablav \potp = \nablav (\rhob\,\potp) \;-\; 
\potp\,\nablav \rhob 
\eep

In order to address the zonal-wind related effects, one considers 
the curl of the Navier-Stokes equation \refp{eq:NSP} where the 
pressure gradient and the conservative part of \eqref{eq:GP} drop out. 
The approximation motivated in the introduction suggest to 
neglect inertia, viscous effects, and the 
Lorentz force contribution:  
\bel{eq:WITHROT}
2\Omega\;\frac{\partial}{\partial z}\left(\rhob\,U_\phi\right) = 
\uvp\cdot\left(\curl \left[\;\rhop\,\nablav\potb_e 
- \potp\,\nablav\rhob\;\right]\right)
\eep
The next step is to assume that $\psi_\Omega$ 
can be neglected in comparison to the background 
gravity contribution $\potb$, as discussed in the 
introduction. 
The background state then becomes spherically symmetric 
and \eqnref{eq:WITHROT} simplifies to 
\bel{eq:TGWE}
2\Omega\;\frac{\partial}{\partial z} \left(\rhob\,U_\phi\right) = 
\frac{1}{r}\,\frac{\partial}{\partial \theta}
\left(\;\rhop\,\frac{\partial}{\partial r}\potb - 
\potp\,\frac{\partial}{\partial r}\rhob\;\right)
\eep
This is the thermo-gravitational wind equation (TGWE) 
solved for a given $U_\phi$ for example 
by \citet{Zhang2015} or \citet{Kong2018}.  
The equation assumes the form of 
a classical thermal wind equation (TWE) when neglecting the 
\DSG, $\rhob\nablav\potp$, or more precisely its 
non-conservative contribution.

Integrating \eqnref{eq:TGWE} in latitude, 
dividing by background gravity $\gvb=-\partial \potb / \partial r$, 
and using \eqnref{eq:potp} finally yields an equation
that connects the perturbation in the gravity potential to
the $z$-gradient of the zonal winds:
\bel{eq:TGWEpot}
\left( \nabla^2 + \mu \right) 
\potp =  4\pi\,\gravc\;\rhoU
\eec
with 
\bel{eq:c}
\mu(r) = 4\pi\,\gravc\,\nablav\rhob\,\big/\,\gb
\eec
and the dynamic density perturbation 
\bel{eq:rhoU}
\rhoU(r,\theta) = \frac{2\Omega r}{\gb}\;\int_0^\theta\,d \hat{\theta}  
\;\frac{\partial}{\partial z}\,\left(\rhob\,U_\phi\right)
\eec
as a source term. Note that $\rhoU$ is an
auxiliary variable different from $\rhop$. 
We will refer to $\mu(r)$ as the \DSG\ coefficient. 

This second order differential equation must be supplemented 
by boundary conditions. Solving for solutions 
in a full sphere, we demand that $\potp$ vanishes at $r=0$. 
Outside of the source, the solutions must 
obey
\bel{eq:BC0}
\nabla^2\potp=0
\eep 
A respective matching condition at the outer radius $R$ yields the
second boundary condition that we provide further below.

Because $\rhoU$ is axisymmetric, we will only consider 
axisymmetric solutions. The integration in latitude 
means that \eqnref{eq:TGWEpot} is only 
determined up to an arbitrary function of radius. 
This function could only contribute to the spherical symmetric 
gravity contribution which, outside of the 
planet, is determined by its total mass 
and thus carries no information on the dynamics.

The case of the TWE is easy to deal with. Neglecting 
the \DSG\ implies $\rhoU=\rhop$ and one simply 
has to solve the classical Poisson equation \refp{eq:potp}. 
The case of the TGWE is more complicated. 
Using \eqnref{eq:VC} and \eqnref{eq:GM} transforms   
the TGWE into the complicated integro-differential 
equation for $\rhop$ derived by \citet{Zhang2015} 
and the Possion equation for $\potp$ is then solved in 
a second step. 
Their solution is cumbersome and numerically time-consuming. 
We avoid this complication by directly solving the
inhomogeneous Helmholtz-type \eqnref{eq:TGWEpot} 
to obtain $\psi '$. 
The true density perturbation can be recovered by
\bel{eq:rhopp}
\rhop = \rhoU - \frac{\mu}{4\pi\gravc}\;\potp
\eec
which is obtained from \eqnref{eq:potp} and \eqnref{eq:TGWEpot}. 
We note that $\rho^U$ is identical to the 'effective density'
that had been introduced by \citet{Braginsky1995} in the 
context of geodynamo equations. They showed that using 
this variable is an elegant way of dealing with self-gravity, 
which greatly simplifies that system of equations to be solved.   

\begin{figure}
\centering
\includegraphics[width=0.7\columnwidth]{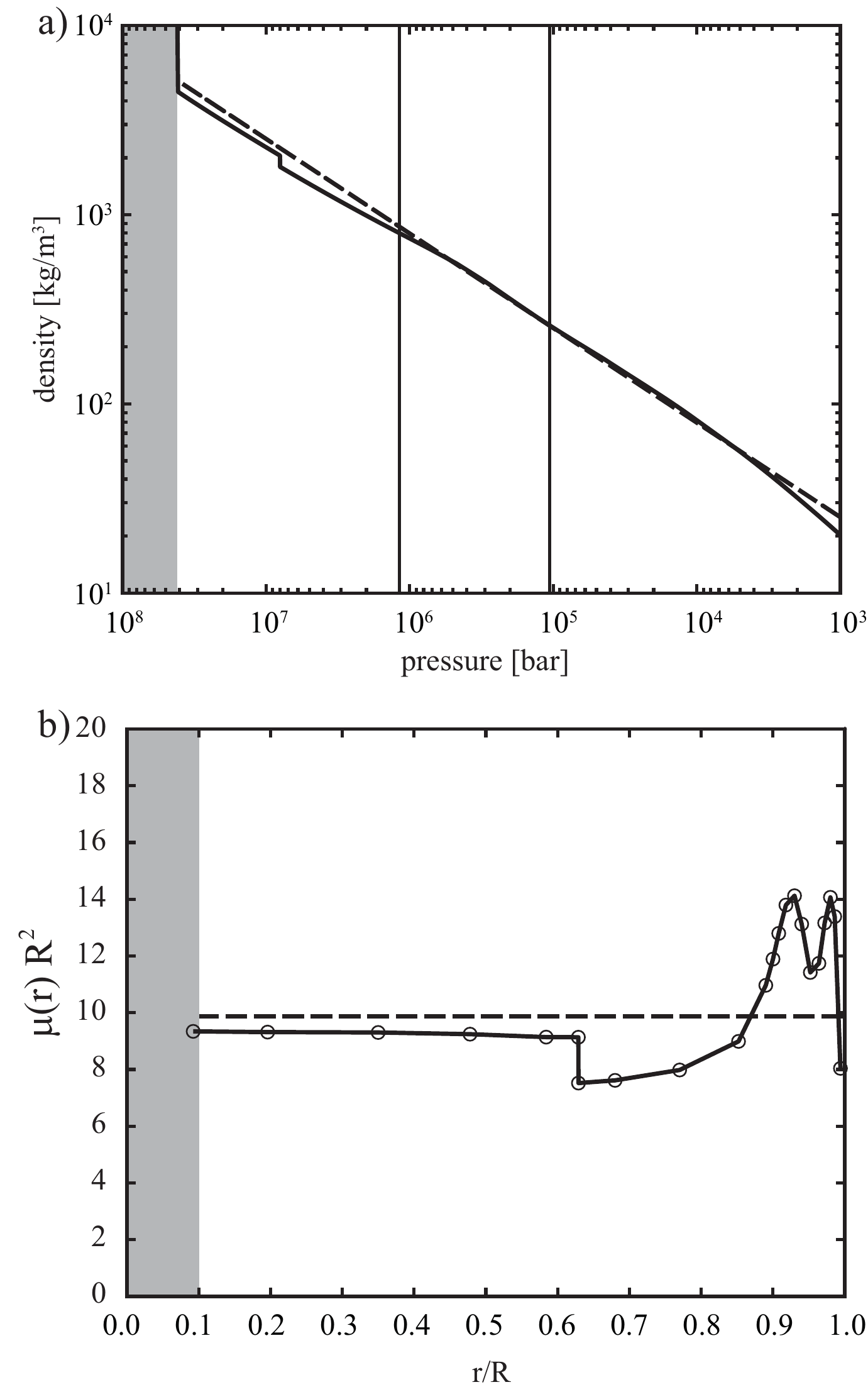}
\caption{Panel a) shows pressure versus density (solid line) 
for the Jupiter model by \citet{Nettelmann2012} and \citet{French2012}
and a polytrope of index unity (dashed line).  
The double logarithmic plot highlights that 
this polytrope, i.e.~$p\sim\rho^{2}$, provides 
a decent approximation. 
The Jupiter model by \citet{Nettelmann2012} and \citet{French2012} 
is a three layer model with a rocky 
core that occupies the inner $10$\% in radius and two gaseous envelopes, 
above and below $0.625\,R$, which 
differ in the metallicity (fraction of elements heavier then 
helium). 
Panel b) compares the normalized \DSG\ profile $\mu(r)\,R^2$ 
(solid line) suggested by the ab-initio data points by 
\citet{French2012} (circles) with the constant value 
$\pi^2$ expected for the polytrope (dashed line).  
}
\label{fig:cr}
\end{figure}

What would be a realistic \DSG\ coefficient $\mu$?
Typical textbook density and pressure profiles 
consider polytropes with index unity. 
They not only seem to provide reasonable 
approximations for Jupiter's interior, as is illustrated in
\figref{fig:cr}, but also yield an
analytical expression of the background density 
and gravity. 
The former is given by 
\bel{eq:rhoPG}
 \rhob(r) =  \rhob_c\,\frac{\sin{\chi}}{\chi}
\eec
where $\rho_c$ is the density at $r=0$, and $\chi$ 
a rescaled radius: 
\bel{eq:chi} 
\chi= \pi\;\frac{r}{R}\,\frac{\rho_c-\rho(R)}{\rho_c}
\eep
The gravity profile is then 
\bel{eq:gp} 
  \gb(r) = - 4\pi\,\gravc\;
  \frac{\rho_c^2}{\rho_c-\rho(R)}\;\frac{R}{\pi}\;
  \frac{\chi \cos{\chi}-\sin{\chi}}{\chi^2}
\ee
and the \DSG\ coefficient becomes constant:
\bel{eq:cZ}
 \mu(r)= \frac{\pi^2}{R^2}\;
 \left(\frac{\rho_c-\rho(R)}{\rho_c}\right)^2
 \approx \frac{\pi^2}{R^2}
\eep
Panel a) of \figref{fig:cr} compares the pressure 
profile in the Jupiter model by \citet{Nettelmann2012} and
\citet{French2012} with a polytrope with index unity,
illustrating that this indeed provides a good approximation.

More generally, for an adiabatic background state, the 
density gradient can be written in terms
of a pressure gradient:
\bel{eq:DG}
\nablav \rhob = \beta_S\,\rhob\; \nablav \pb
\eec
with 
\bel{eq:betas}
\beta_S = \frac{1}{\rhob}\left(\frac{\partial \rho}{\partial p} \right)_S 
\ee
being the compressibility at constant entropy. 
Combining \eqnref{eq:DG} and \eqnref{eq:HS} shows that 
the gradient in the background density is given by
\bel{eq:drhob}
\frac{\partial}{\partial r} \rhob = \beta_S\,\rhob^2\,\gb
\eep
The \DSG\ coefficient is thus given by
\bel{eq:cA}
 \mu(r) = 4\pi \gravc\; \beta_S\,\rhob^2
\eep 

Panel b) of \figref{fig:cr} compares the constant 
expression \refp{eq:cZ} for the index-unity polytrope (dashed line) 
with the profile \refp{eq:cA} based on ab-initio 
equation-of-state simulations and pre-Juno 
gravity data \citep{French2012}. 
Considering the strong variation of other thermodynamic 
quantities, the $\mu(r)$ variations remain remarkable small.
In the lower layer $r<0.25 R$, $\mu(r)$ is nearly 
constant and close to $\pi^2/R^2$. In the outer envelope   
$r>0.85\,R$, $\mu$ becomes more variable, reaching 
amplitudes $40$\% larger than $\pi^2/R^2$.
A constant $\mu$ value thus seem to provide 
a decent approximation and  will considerably 
ease the task of solving the inhomogeneous Helmholtz equation,
as we will discuss in \secref{sec:method}. 

\section{Solving Poisson and Inhomogeneous\\ Helmholtz Equations}
\label{sec:method}

We start with briefly recapitulating the Green's function method
for solving the Poisson equation in \secref{sec:classic}. 
\Secref{sec:Helm} then discusses the adapted approach
for solving the inhomogeneous Helmholtz equation with constant 
\DSG\ coefficient $\mu$. 
The involved methods represent textbook knowledge, but  
their application to the specific gravity problem is new 
however, we nevertheless discuss them in some detail.

\subsection{The Classic Green's-Function Solution}
\label{sec:classic}
A common way of solving the Poisson equation \refp{eq:poiss} 
is the Green's function method. 
The respective Green's function $\GF$ is defined by 
\bel{eq:GF}
\nabla^2 \GF(\rv,\other{\rv}) = \delta ( \rv-\other{\rv} ) 
\eec
where vectors $\rv$ and $\other{\rv}$ denote the location of potential 
and density, respectively.  The Green's function also has to 
fulfill the same boundary conditions as the gravity potential. 
The solution is then given by the integral
\bel{eq:GS}
\pot(\rv) =  4\pi\,\gravc \;\int\,d\,\other{V}\; \GF(\rv,\other{\rv})\,\rho(\other{\rv})
\eec
where 
\bel{eq:dV}
\int\,d\,\other{V} = \int_{0}^{R}\,d\,\other{r}\;{\other{r}}^2\;
\int_0^\pi\,d\,\other{\phi}\;\int_0^{2\pi}\,d\,\other{\theta} 
\sin{\other{\theta}}
\ee
denotes the integration over the spherical volume. 

The classical Green's function for the Poisson problem is given by
\bel{eq:GFC}
 \GF(\rv,\other{\rv}) = - 1 \big/\,
 \left(4\pi\left|\rv-\other{\rv}\right|\right)
\eec
but of more practical use is the representation 
where $\Gamma$ is expanded in eigenfunctions of 
the Laplace operator. 
Since the Legendre polynomials 
are eigenfunctions of the horizontal
part of the Laplace operator, they are a natural choice to 
describe the latitudinal dependence:  
\bel{eq:LH}
\nabla^2\;f(r)\;P_\ell(\theta) \;=\; 
\left(\;\frac{\partial^2}{\partial r^2}\,+\,
\frac{2}{r}\frac{\partial}{\partial r}\,-\,
\frac{\ell(\ell+1)}{r^2}\;\right)\;
f(r)\;P_\ell(\theta) 
\eep
The Schmitt normalization assumed here means that 
\bel{eq:ON}
\int_0^\pi\,d\theta \sin\theta\;
P_{\ell}(\theta)\;P_{\ell^\prime}(\theta)\;=\;\frac{2}{2\ell+1}\;
\delta_{\ell \ell^\prime}
\eep  
The two possibilities for the radial function are 
$f_\ell (r)=r^\ell$ and $f_\ell (r)=r^{-(\ell+1)}$.
The expanded Green's function then reads   
\bel{eq:GFC2}
\GF(\rv,\other{\rv}) = - \frac{1}{4\pi}
 \sum_{\ell=0}^{\infty}\;\frac{{r_<}^\ell}{r_>^{\ell+1}} \;
P_{\ell}(\theta)\;P_{\ell}(\other{\theta})
\eec
where $r_>$ ($r_<$) denotes that larger (smaller) of 
the two radii $r$ and $\other{r}$. 
The matching condition to the field for $r>R$ reduces 
to the mixed boundary condition
\bel{eq:BC}
 \frac{\partial}{\partial r}\;f_\ell(r) = - \frac{(\ell+1)}{R}\;f_\ell(r)
\eec
which is obviously fulfilled by the radial ansatz functions
and thus by the Green's function. 

Plugging the Green's function into \eqnref{eq:GS} 
then shows that the potential field for $r>R$ is given by 
\bel{eq:VE}
\pot(\rv) = \sum_{\ell=0}^{\infty}\; 
\pot_{\ell}\;\left( \frac{R}{r} \right)^{\ell+1}\;P_{\ell}(\theta) 
\eec
with the expansion coefficients 
\bel{eq:VEC}
\pot_{\ell} = -\frac{\gravc}{4\pi R}\;
\int\,d \other{V}\;\left(\frac{\other{r}}{R}\right)^\ell\;
\rho(\other{\rv})\,P_{\ell}(\other{\theta})
\eep
This is equivalent to the differently normalized 
classical expansion \eqnref{eq:VC} and \eqnref{eq:GM}.

The same solution applies to $\potp$ when replacing $\rho$ by 
$\rhop$. Should the impact of \DSG\ $\mu$ be negligible, 
we could simply use $\rhop\approx\rhoU$, 
an approach generally followed by one group of authors 
mentioned in the introduction 
\citep{Kaspi2016,Galanti2017,Kaspi2018,Iess2019,Galanti2019}. 

\subsection{Solving the Inhomogeneous Helmholtz equation}
\label{sec:Helm}
 
For constant $\mu(r)=K^2$, the modified potential field 
equation becomes an inhomogeneous Helmholtz equation 
\bel{eq:Helm}
\left(  \nabla^2 + K^2 \right)\;\potp= 4\pi\,\gravc\;\rhoU 
\eep
The respective Green's function is now defined by 
\bel{eq:GFD}
\left(\,\nabla^2 + K^2\,\right) \GF = \delta( \rv-\other{\rv} )
\eep
and has to fulfill the boundary conditions. 

Like for the classical Green's function solution 
discussed in \secref{sec:classic}, we are looking for a solution 
in terms of orthonormal functions. 
While Legendre polynomial can once more be used 
for the horizontal dependencies, the radial functions 
have to be different. 
We will rely on eigenfunctions $f_\ell(r) P_\ell(\theta)$ 
of the Laplace operator 
where the $f_\ell(r)$ fulfill the boundary conditions. 

An orthonormal set of such radial functions 
can be constructed from spherical Bessel functions 
\citep{Abramowitz1984},
which solve the differential equation  
\bel{eq:SBFE0}
\left( \frac{\partial^2}{\partial r^2} + 
\frac{2}{r}\frac{\partial}{\partial r} - 
 \frac{\ell (\ell+1)}{r^2}\;+\;1\right) j_\ell(r) = 0
\eep
We only use the spherical Bessel functions of the first kind, $j_\ell$, 
with $\ell>0$ that all vanish at $r=0$. 
Spherical Bessel functions of the second kind diverge at the origin,
while $j_0(r=0)=1$. 
Simple rescaling of the argument yields eigenfunctions 
of the Laplace operator: 
\bel{eq:SBFE}
\nabla^2 \;j_\ell(\kln r)\;P_{\ell}(\theta)=  \lambda\; 
j_\ell(\kln r)\;P_{\ell}(\phi)
\eec
with eigenvalues
\bel{eq:EVs}
 \lambda=-\kln^2 
\eep

\begin{figure}
\centering
\includegraphics[width=0.7\columnwidth]{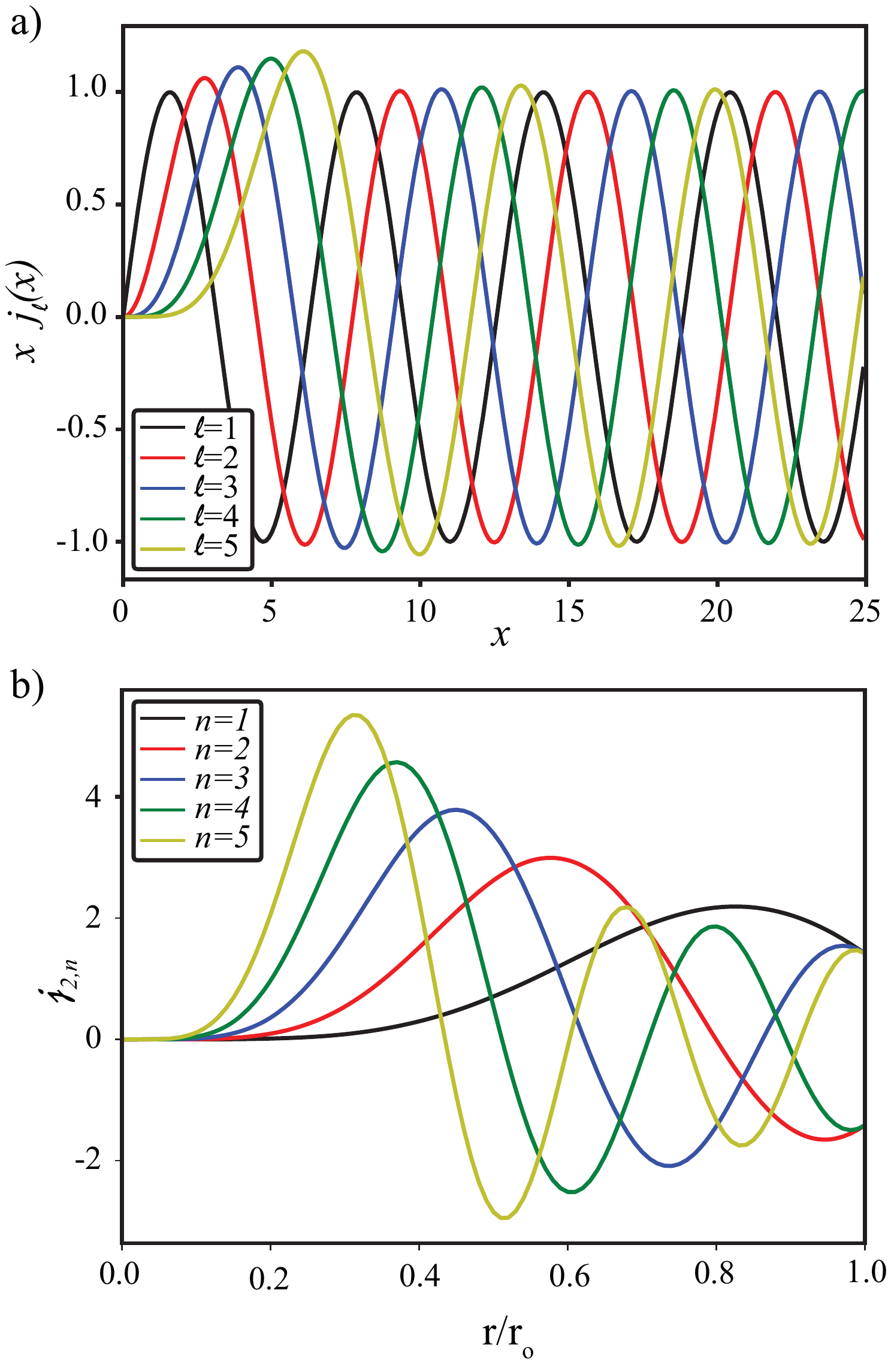}
\caption{
Panel a) shows the first five spherical Bessel functions
of the first kind. Panel b) shows the first orthonormal 
normalized functions $\cjln$ for degree $\ell=2$. 
}
\label{fig:js}
\end{figure}

The different $\kln$ are chosen so that $j_\ell(\kln R)$  
fulfills the boundary condition \refp{eq:BC}. 
Because of recurrence relation \refp{eq:RC1} 
(see \appref{sec:RR}), this condition reduces to 
\bel{eq:BC2}
 j_{\ell-1} ( \kln R ) =0
\eec
which means that the $\kln$ are the roots of 
$j_{\ell-1}(x)$ divided by the outer boundary radius $R$. 
We start the numbering at the smallest  
root larger than zero so that 
$0 < k_{\ell 1} < k_{\ell 2} < k_{\ell 3} < ...$.
Panel (a) of \figref{fig:js} illustrates the spherical Bessel functions $j_\ell$
for different degrees $\ell$. 
\Tabref{tab:kln} list the first five roots for 
$\ell\le 5$.

\begin{table}
\centering
\begin{tabular}{c|*{5}{c}}
$\ell$ \big/ $n$ &1&2&3&4&5 \\
\hline
 1 & 1 & 2 & 3 & 4 & 5 \\
 2 & 1.4303 & 2.4590 & 3.4709 & 4.4774 & 5.4815 \\
 3 & 1.8346 & 2.8950 & 3.9225 & 4.9384 & 5.9489 \\
 4 & 2.2243 & 3.3159 & 4.3602 & 5.3870 & 6.4050 \\
 5 & 2.6046 & 3.7258 & 4.7873 & 5.8255 & 6.8518 \\
\end{tabular}
\caption{List of $\kln R/\pi$. 
The $\kln R$ are the roots of $j_{\ell -1}$.}
\label{tab:kln}
\end{table}

Since the Laplace operator is hermitian (adjoint) and 
our radial ansatz functions fulfill the boundary conditions,  
the eigenvalues are real and 
the eigenfunctions for different eigenvalues are 
orthogonal. For completeness, we include this
textbook knowledge is \appref{sec:ortho}.  
The orthonormality condition thus reads  
\bel{eq:KO}
 N_{\ell n}\;N_{\ell n^\prime}\;\int\,d r\;r^2\;j_\ell(\kln r)\; 
 j_\ell(k_{\ell n^\prime} r) = \delta_{n,n^\prime}
\eec 
where the $N_{\ell n}$ 
are normalization constants derived analytically in \secref{sec:norm}: 
\bel{eq:norm}
N_{\ell n}= \left(\frac{2}{R^3 j_\ell^2(\kln R)}\right)^{1/2}
\eep
Panel (b) of \figref{fig:js} shows the first five 
normalized functions,
\bel{eq:cj}
 \cjln(r)=N_{\ell n}\; j_\ell(\kln r)
\eec 
for $\ell=2$. 

We can now expand the potential field perturbation in Legendre 
polynomials and the new orthonormal radial functions: 
\bel{eq:PE}
\potp(\rv) = \sum_{n=1}^\infty\,\sum_{\ell=1}^{\infty}\;
    \potp_{\ell n}\;\cjln(r)\,P_{\ell}(\theta)
\eep
Using this expansion in \eqnref{eq:Helm}, multiplying with 
the ansatz functions $\cjln(\other{r}) P_{\ell}(\other{\theta})$ 
and integrating over the volume yields a spectral equation  
for the expansion coefficients: 
\bel{eq:SE}
   \frac{4\pi\,\left( \kln^2 - K^2 \right)}{(2\ell+1)}\;
   \potp_{\ell n} =  
   - 4\pi\;\gravc\;\int\,d\,V\;\rhoU(\rv)\;
   \cjln\;(\other{r})\,P_{\ell}(\other{\theta})
\eep
The coefficients are thus simply given by  
\bel{eq:EE2}
    \potp_{\ell n} = 
    -\frac{\gravc\,(2\ell+1)}{\kln^2 -K^2} 
    \;\int\,d\,\other{V}\;\rhoU(\other{\rv})\;
   \cjln(\other{r})\;P_{\ell}(\other{\theta}) 
\eep

A comparison with \eqnref{eq:GFC2} shows that the Green's function
for the inhomogeneous Helmholz equation is then 
\bel{eq:GFH}
   \GF(\rv,\other{\rv}) = - \frac{1}{4\pi}
   \sum_{n=1}^\infty\,\sum_{\ell=1}^{\infty}\;
   \frac{(2\ell+1)}{\kln^2- K^2}\;
   \cjln(r)\;\cjln(\other{r})\;P_{\ell}(\theta) 
   P_{\ell}(\other{\theta})
\eep

The potential field for $r>R$ has to decay 
like $\left(R/r\right)^{\ell +1}$.
The respective solution is thus given by
\bel{eq:RSo}
\potp(\rv) = \sum_{\ell=1}^{\infty}\; 
\potp_{\ell}(R)\;\left(\frac{R}{r}\right)^{\ell+1}\;P_{\ell}(\theta)  
\eec
with
\bel{eq:potlm}
 \potp_{\ell}(R) = \sum_{n=1}^{\infty}\; 
 \potp_{\ell n}\;\cjln(R)
\eep

As expected, this solution is identical to 
the classical result \refp{eq:VE} for $K^2=0$. 
We show this analytically in \secref{sec:ident}.

\section{Illustrative Examples}
\label{sec:solution}

We can easily convince ourselves that \eqnref{eq:PE}
with coefficients \refp{eq:EE2}  
provides a correct solution when 
assuming that the source is given by only one ansatz function:
\bel{eq:ex1}
  \rhoU=\cjln(r)\;P_\ell(\theta)
\eep
Only the respective potential field coefficient thus has 
to be considered and the solution for $r<R$ is 
\bel{eq:ex1pot}
   \potp(\rv) =  - \frac{\gravc\,(2\ell+1)}{\kln^2-K^2}\;
   \cjln(r)\;P_\ell(\theta)
\eep
Solving for a more general source thus boils down to the 
question: How well can $\rhoU$ be expanded in the ansatz functions? 

A special situation arises when $K^2=\kln^2$.
For the polytropic density distribution with
polytropic index unity, this happens 
for $\ell=1$ and $n=1$ where $K=k_{1,1}=\pi/R$.
The two non-conservative buoyancy terms
then cancel exactly,
\bel{eq:CAN}
  \rhop  \nablav \potb - \potp \nablav \rhob = 0
\eec
because of matching radial functions in the 
background profiles and the primed perturbations. 
Nothing is left to balance the respective left
hand side of the simplified dynamic equation 
\refp{eq:TGWE} or the related 
$\rhoU$ contributions in \refp{eq:TGWEpot}. 
The respective potential field perturbation thus 
decouples from the simplified dynamical equation. 

Even when $K$ is not identical but close to $k_{1,1}$, the dynamic 
equation requires an unrealistically large potential
field perturbation and the precise value of $K$ 
would have an enormous effect. 
It thus seems a good idea to generally avoid these 
resonance conditions and we will simply not 
interpret respective $\potp_{1,1}$ contributions. 
Since the $\ell=1$ gravity contribution generally 
vanishes due to the choice of origin $r=0$, these
considerations are of little practical use.


Partial integration of the dynamic density perturbation yields 
\begin{multline}
\label{eq:rhoUI}
\rhoU = \frac{2\Omega}{\gb}\;
\left( \rhob\,\sin{\theta}\,U_\phi + 
r \frac{\partial\rhob}{\partial r}\;\int_0^\pi\,d\theta\;\cos\theta\,U_\phi\right.
 + \\ \left.
r\,\rhob\;\int_0^\pi\,d\theta\;\cos\theta\,\frac{\partial U_\phi}{\partial r}
\;\right)\;\;.
\end{multline}
While latitude-dependence is this purely determined 
by the zonal flow, $\rhob$, $U_\phi$ and 
their radial derivatives influence the 
radial profile of $\rhoU$. 

Since the expansion of the latitude-dependence 
in Legendre polynomials is not specific to solutions 
with or without \DSG, we concentrate 
on discussing the expansion in radius. 
The steep radial gradients in density and zonal flows 
characteristic for gas planets may prove 
challenging here. 

Choosing a truncation $N$ for the radial 
expansion defines the numerical representation of $\rhoU$: 
\bel{eq:rhoUE}
 \rho^{U\!N}_{\ell}(r) = \sum_{n=1}^{N}\; 
     \rho^{U}_{\ell n }\;\cjln(r)
\eec
with
\bel{eq:rhoUEln}
  \rho^{U}_{\ell n} = 
  \int_0^{R}\,d\,r\;r^2\;
  \rhoU_\ell(r)\;\cjln(r)
\eec
and
\bel{eq:rhoUl}
  \rho^{U}_{\ell}(r) = 
  \int_0^{\pi}\,d\,\theta\;
  \sin{\theta}\;\rhoU(r,\theta)\;\Plt
\eep
The quality of the representation is quantified by the 
misfit 
\bel{eq:M}
    D(N) = \frac{ \int_0^{r_o}\,d\,r\;r^2\;
    \left[\,\rho^{U\!N}_{\ell}(r) - \rhoU_{\ell}(r)\,\right]^2}
    { \int_0^{r_o}\,d\,r\;r^2\;{\rhoU_{\ell}}^2(r)}
\eep

\begin{table}
\centering
\begin{tabular}{c|*{4}{c}}
 N    & \multicolumn{2}{c}{h=0.143} & \multicolumn{2}{c}{h=1.143} \\
     & TWE & TGWE & TWE & TWGE  \\
\hline
 10 &$3.133\tp{-5}$&$4.961\tp{-5}$&$0.8419\tp{-4}$&$1.489\tp{-4}$ \\
 20 &$3.165\tp{-5}$&$4.992\tp{-5}$&$0.8433\tp{-4}$&$1.491\tp{-4}$ \\
 40 &$3.169\tp{-5}$&$4.997\tp{-5}$&$0.8435\tp{-4}$&$1.491\tp{-4}$ \\
 60 &$3.170\tp{-5}$&$4.998\tp{-5}$&$0.8435\tp{-4}$&$1.491\tp{-4}$ \\
100 &$3.170\tp{-5}$&$4.998\tp{-5}$&$0.8435\tp{-4}$&$1.491\tp{-4}$ \\
\hline
Z2015&$3.17\tp{-5}$&$5.00\tp{-5}$&$0.874\tp{-4}$&$1.553\tp{-4}$
\end{tabular}
\caption{Gravity harmonic $J_2$ for the equatorially symmetric 
test case suggested by \citet{Zhang2015}. 
Column 2 and 3 list TWE and TGWE results for the slower 
decaying flow with $h=0.143$. Column 4 and 5 list respective values
for the faster decaying case $h=1.143$ also illustrated in \figref{fig:Zhang}. 
The last line lists the values published by \citet{Zhang2015}.} 
\label{tab:Zhang}
\end{table}

We start with exploring a test case suggested by \citet{Zhang2015}.
They assume the polytrope index unity density profile 
\refp{eq:rhoPG} and a zonal flow defined by 
\bel{eq:UZ} 
U_\phi = U_0\;f_1(r)\;\sin^2{\theta}
\ee
with amplitude $U_0=R\Omega/100$ 
and radial dependence 
\bel{eq:f1}
 f_1(r) = \left(\frac{r}{R}\right)^2\;\exp{\left(-\frac{1}{h}\frac{R-r}{R}\right)}
\eep
Jupiter values used to define flow and gravity are 
$R=6.9894\tp{7}\,$m, $\Omega=1.759\tp{-4}\,$s$^{-1}$, and $M=1.898\tp{27}\,$kg. 
Two relative decay scale heights $h=0.143$ and $h=1.143$ are explored. 
The flow yields $\ell=0$ and $\ell=2$ gravity perturbations, 
but since the former would be nonphysical in a real gravity 
problem we only consider the latter.  
\Tabref{tab:Zhang} compares the respective $J_2$ coefficients published 
by \citet{Zhang2015} with values for different truncations $N$. 
While the results for $h=0.143$ exactly match those of 
\citet{Zhang2015}, those for $h=1.143$ already differ in the second figure. 
We attribute this to convergence problems reported by \citet{Zhang2015}. 

\begin{figure}
\centering
\includegraphics[width=1.0\columnwidth]{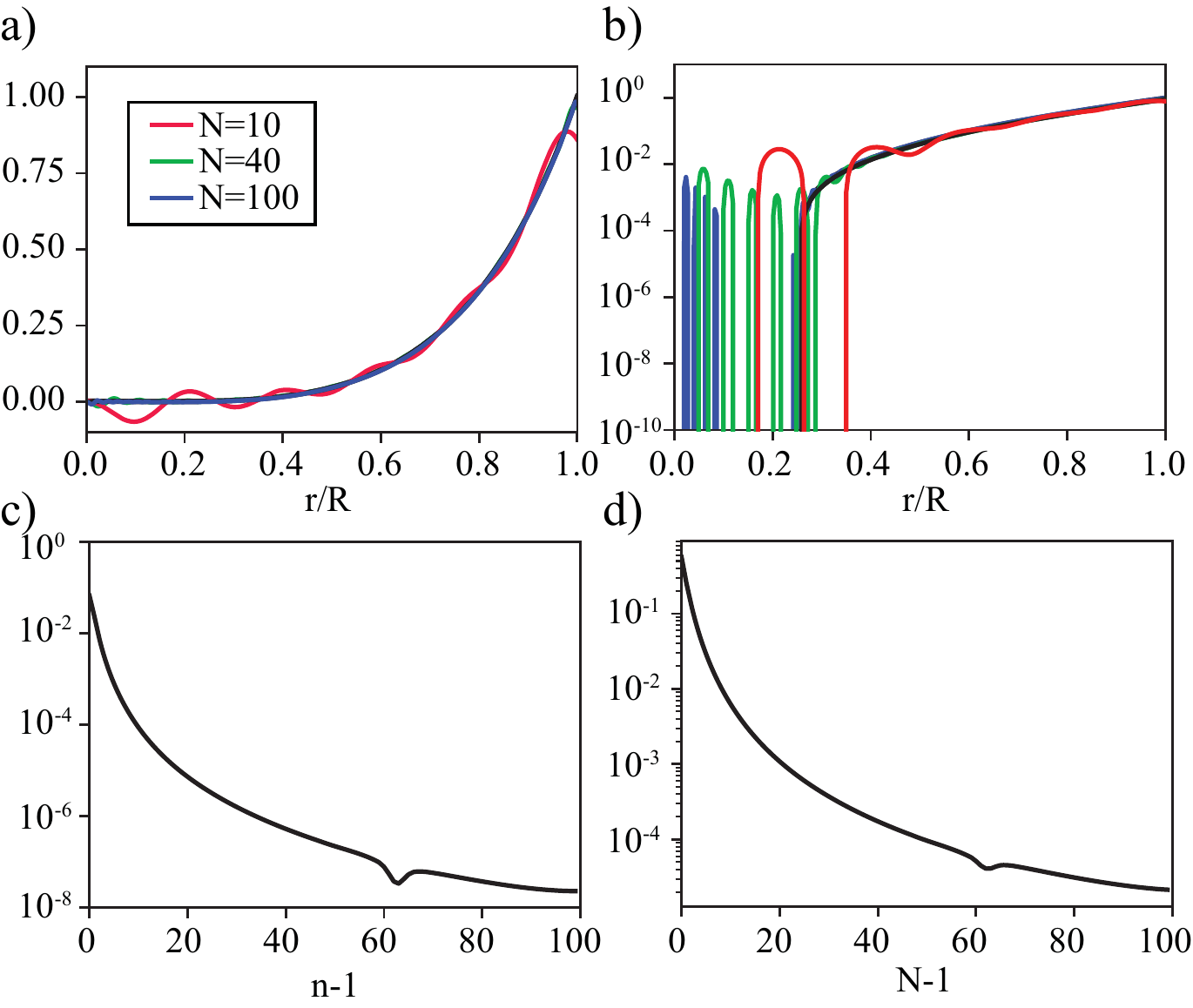}
\caption{Expansion of the function $f_1(r)$ with $h=1.143$ into the $\cjln$ for 
$\ell=2$. Panel a) compares the normalized function with representations for truncations $N=10$, $40$, and $100$. 
Panel b) shows the same 
in a logarithmic plot. Panel c) shows the spectrum for $N=101$ 
and panel d) the misfit $D(N)$.}
\label{fig:Zhang}
\end{figure} 

The well behaved convergence for the expansion of $f_1(r)$ is 
documented in \tabref{tab:Zhang} and illustrated in \figref{fig:Zhang}. 
Panel a) and b) demonstrate that the function is already almost perfectly 
represented with a truncation of $N=40$. 
Small differences tend to remain close to the outer boundary and 
at small radii due to the specific properties  
of the $\cjln$.  
Spectrum and misfit $M$, depicted in panels c) and d) respectively, 
decay continuously with truncation but with a slower rate at 
higher degrees because of the difficulties in exactly 
capturing the vanishing values for $r\rightarrow0$.  

\begin{figure}
\centering
\includegraphics[width=1.0\columnwidth]{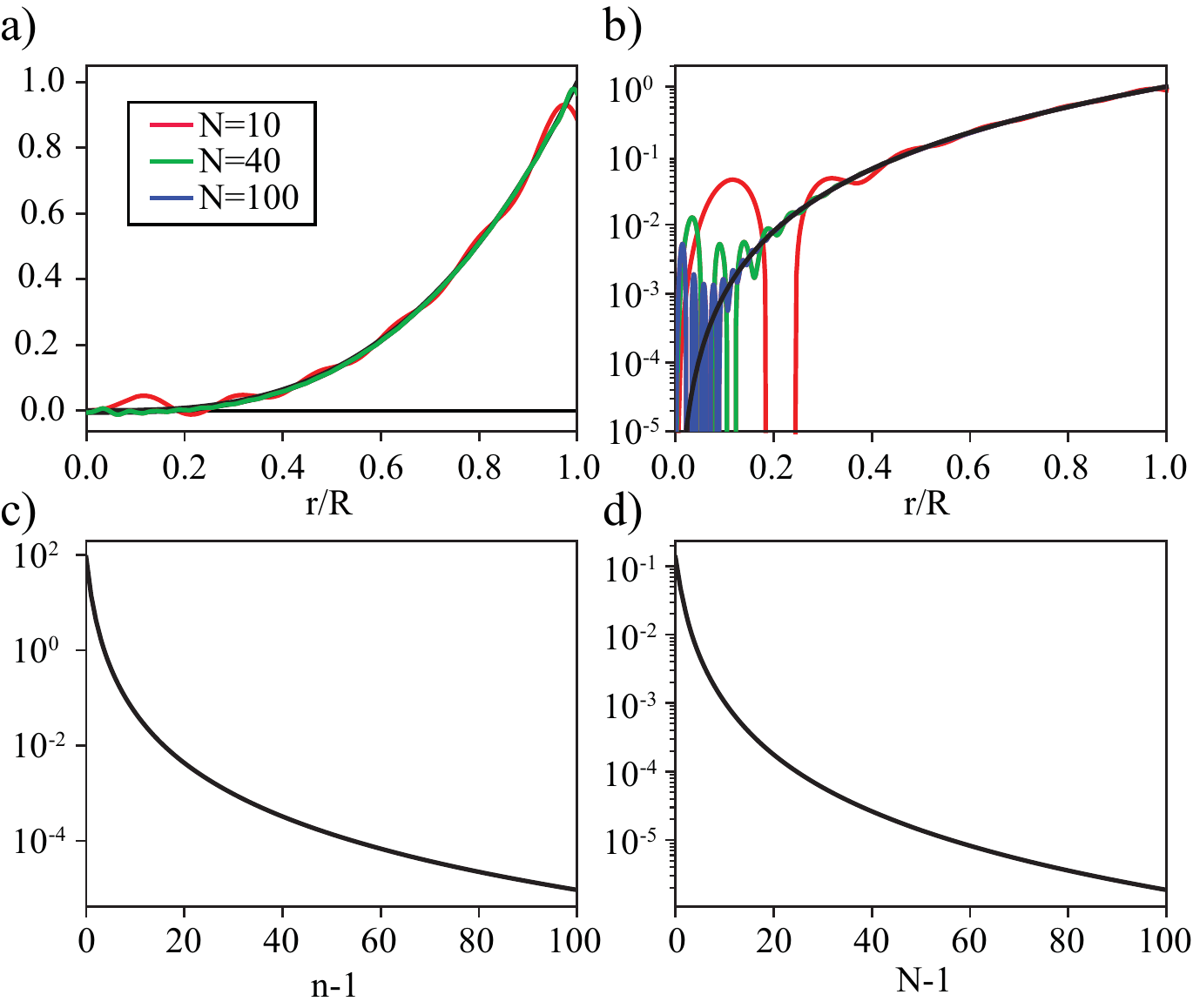}
\caption{Expansion of the function $f_2=r^3$ for $R=10$ into the $\cjln$ 
for $\ell=3$. Panel a) compares the normalized function with 
representations for truncations $N=10$, $40$, and $100$. 
Panel b) shows the same 
in a logarithmic plot. Panel c) shows the spectrum for $N=101$ 
and panel d) the misfit $D(N)$.}
\label{fig:testrl}
\end{figure}

As a second example we explore the function  
\bel{eq:rl}
       f_2(r) = r^\ell
\ee
used in the classical potential field solution 
for $K^2=0$. This is an ideal test case, since the 
expansion coefficients are known analytically 
(see \appref{sec:ident}). 
\Figref{fig:testrl} illustrates the quality
of the expansion for $\ell=3$. Panels a) and b) once more 
illustrate the difficulties of representing the 
function at the boundaries. 

The last example is the radial function 
\bel{eq:rhogdU}
 f_3(r) = \frac{\rhob\,r}{\gb}\;\frac{\partial U_\phi(r) }{\partial r}
\ee
that determines the radial dependence of one term in 
$\rhoU$ according to \eqnref{eq:rhoUI}. 
Following the example of \citet{Kong2018}, we assume 
a polytrope of index one and the Gaussian-like flow profile: 
\bel{eq:U}
  U_\phi (r) = \left\{ 
  \begin{array}{ll} 
     \exp{\left( \frac{1}{h}\,\frac{d^2}{D^2 - d^2}\right)} & \mbox{for}\;d\le D \\
     0 & \mbox{for}\;d> D
  \end{array}
  \right.
\eep
where $d=R-r$ is the depth, $D=0.15\,R$ 
is the maximum depth of $\rhoU$, 
and $h=0.22$ determines the decay rate. 

\begin{figure}
\centering
\includegraphics[width=1.0\columnwidth]{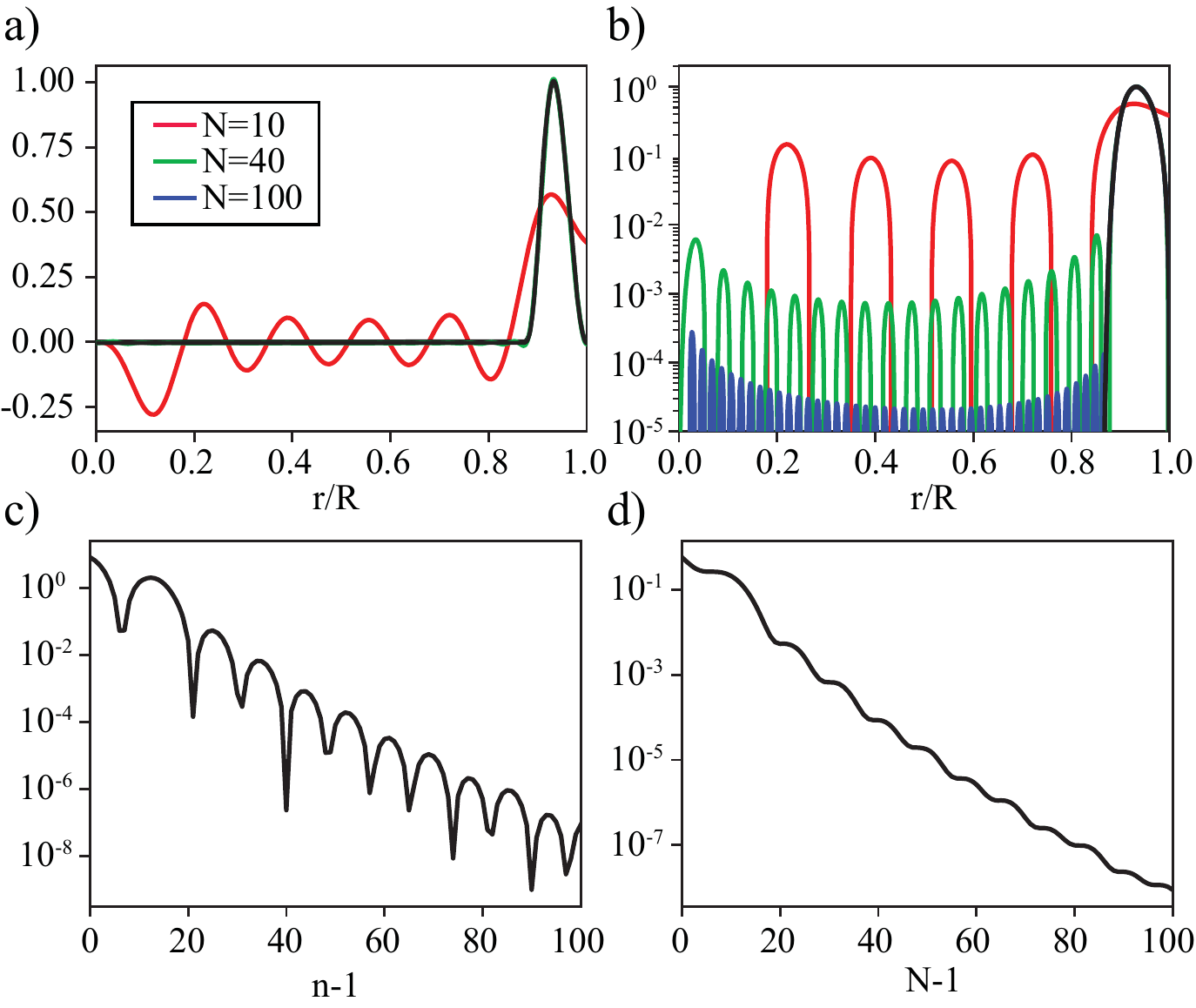}
\caption{Same as \figref{fig:testrl} but for function $f_3(r)$. 
The $\cjln$ for $\ell=3$ have been used.} 
\label{fig:testU1}
\end{figure}

\Figref{fig:testU1} demonstrates 
that the resulting highly localized function is also 
already well represented for a truncation of $N=40$. 
Overall, spectrum and misfit once more decay with growing $N$,
which confirms that there are no principal 
numerical problems with 
expand this demanding function into the $\cjln$. 
The pronounced length scale defined by the width of the function peak 
leads to the local minima in the spectrum where they match 
the distance between the zero intercepts in the $\cjln$. 

\section{Relative Importance of Dynamic Self Gravity}
\label{sec:impact}

The analytical solution shows that the impact of 
the \DSG\ simply depends on the ratio $\kln^2/K^2$.
The relative importance of $K^2$ in the 
inhomogenous Helmholtz equation for a given spherical 
harmonic degree $\ell$ and radial index $n$ 
can be quantified by 
\bel{eq:RIn}
 S_{\ell n} = \frac{ \left(\kln^2-K^2\right)^{-1}}{\kln^{-2}}\;-\;1 
 = \frac{1}{\kln^2/K^2 - 1}
\eep 

\Tabref{tab:imp} lists $S_{\ell n}$ for 
spherical harmonic degrees up to $\ell=30$ 
and $n$ up to $5$, assuming $K=\pi$. 
The values indicate that the \DSG\ should be considered a first 
order effect for $\ell\le4$, reaches the $10$\% level 
at $\ell=5$ or $\ell=6$ and amounts to only about $1$\% 
for $\ell\ge20$. 

\begin{table}
\centering
\begin{tabular}{c|*{5}{c}}
$\ell$ \big/ $n$ &1&2&3&4&5 \\
\hline
 1 &  --- & $3.3\tp{-1}$ & $1.2\tp{-1}$ & $6.7\tp{-2}$ & $4.2\tp{-2}$ \\
 2 & $9.6\tp{-1}$ & $2.0\tp{-1}$ & $9.1\tp{-2}$ & $5.3\tp{-2}$ & $3.4\tp{-2}$ \\
 3 & $4.2\tp{-1}$ & $1.4\tp{-1}$ & $7.0\tp{-2}$ & $4.3\tp{-2}$ & $2.9\tp{-2}$ \\
 4 & $2.5\tp{-1}$ & $1.0\tp{-2}$ & $5.6\tp{-2}$ & $3.6\tp{-2}$ & $2.5\tp{-2}$ \\
 5 & $1.7\tp{-1}$ & $7.8\tp{-2}$ & $4.6\tp{-2}$ & $3.0\tp{-2}$ & $2.2\tp{-2}$ \\
 6 & $1.3\tp{-1}$ & $6.2\tp{-2}$ & $3.8\tp{-2}$ & $2.6\tp{-2}$ & $1.9\tp{-2}$ \\
 8 & $7.8\tp{-2}$ & $4.3\tp{-2}$ & $2.8\tp{-2}$ & $2.0\tp{-2}$ & $1.5\tp{-2}$ \\
10 & $5.3\tp{-2}$ & $3.2\tp{-2}$ & $2.2\tp{-2}$ & $1.6\tp{-2}$ & $1.3\tp{-2}$ \\
14 & $3.0\tp{-2}$ & $2.0\tp{-2}$ & $1.4\tp{-2}$ & $1.1\tp{-2}$ & $8.9\tp{-3}$ \\
20 & $1.6\tp{-2}$ & $1.2\tp{-2}$ & $8.9\tp{-3}$ & $7.2\tp{-3}$ & $6.0\tp{-3}$ \\
30 & $7.9\tp{-3}$ & $6.0\tp{-3}$ & $4.9\tp{-3}$ & $4.1\tp{-3}$ & $3.6\tp{-3}$ \\
\end{tabular}
\caption{Relative importance of \DSG\ measured by $S_{\ell n}$ for 
spherical harmonic degrees up to $\ell=30$ and $n$ up to $5$.} 
\label{tab:imp}
\end{table}

When specifying a source term $\rhoU$, we 
can quantify the relative importance of the 
\DSG\  at each spherical harmonic degree by 
\bel{eq:R}
 S_\ell(N) = \frac{ \sum_{n=1}^{N}\,\cjln(R)\;\rho_{\ell n}^{U}
 \big/\left(\kln^2-K^2\right)}
  { \sum_{n=1}^{N}\,\cjln(R)\;\rho_{\ell n}^{U}
     \big/\kln^2  } \;-\;1
\eep

\begin{figure}
\centering
\includegraphics[width=0.7\columnwidth]{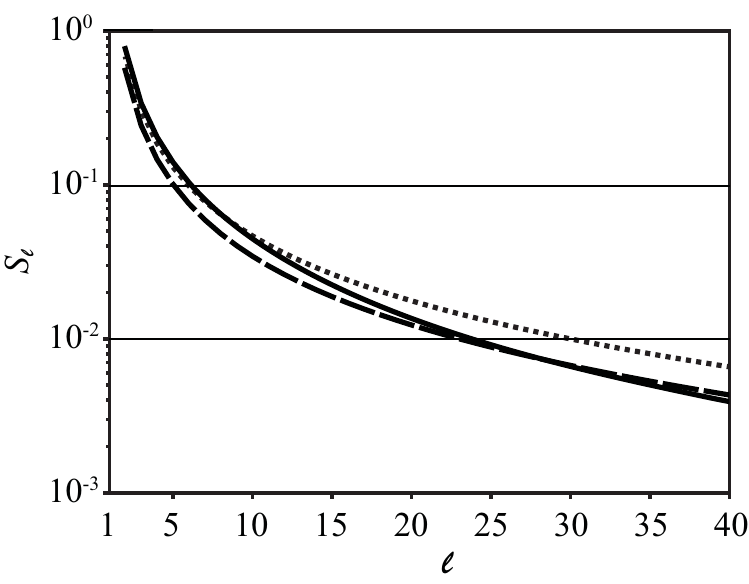}
\caption{Measure $S_\ell(N)$ quantifying the relative importance 
of self gravity at different spherical harmonic 
degrees $\ell$. Line types indicate the different radial 
profiles used for $\rhoU$: 
$f_1=r^5$ (solid), $f_2$ (dotted), and $f_3$ (dashed).}
\label{fig:ratio}
\end{figure} 

\Figref{fig:ratio} compares $S_\ell$ for the 
three radial $\rhoU$ profiles explored in 
\secref{sec:solution}. 
In order to be on the safe side, 
we have used $N=200$. 
Selected values of $S_\ell(200)$ are listed in 
\tabref{tab:ratio}. 
All cases show a similar decay with $\ell$, 
reaching $10$\% relative importance between
$\ell=5$ and $\ell=7$ and $1$\% between
$\ell=22$ and $\ell=30$. 
At least for degrees $\ell>20$, the 
specific radial profile hardly seems to matter. 
Because $n=1$ contributions are always 
significant, the respective ratio \refp{eq:RIn} listed 
in \tabref{tab:imp} already provides a decent estimate of 
the relative importance for the \DSG. 

\begin{table}
\centering
\begin{tabular}{c|*{4}{c}}
$\ell$  & $f_2=r^5$ &  $f_3$  \\
\hline
 2 &$7.6\tp{-1}$ & $6.8\tp{-1}$\\
 3 &$3.4\tp{-1}$ & $3.0\tp{-1}$\\
 4 &$2.0\tp{-1}$ & $1.9\tp{-1}$\\
 5 &$1.4\tp{-1}$ & $1.3\tp{-1}$\\
 6 &$1.0\tp{-1}$ & $1.0\tp{-1}$\\
 8 &$6.4\tp{-2}$ & $6.4\tp{-2}$\\
10 &$4.4\tp{-2}$ & $4.7\tp{-2}$\\
14 &$2.5\tp{-2}$ & $2.9\tp{-2}$\\
20 &$1.3\tp{-2}$ & $1.8\tp{-2}$\\
30 &$6.5\tp{-3}$ & $1.0\tp{-2}$\\
\end{tabular}
\caption{Relative importance of the \DSG\ measured
by $S_{\ell}$ for two different radial functions.
A radial truncation of $N=200$ has been used.}
\label{tab:ratio}
\end{table}

\section{Discussion and Conclusion}
\label{sec:discussion}

The dominant balance between the Coriolis force 
and buoyancy terms in the azimuthal component 
of the vorticity equation establishes a connection
between zonal flows and gravity. 
Simple manipulations lead to what has been called the 
thermo-gravitational wind equation (TGWE) by \citet{Zhang2015}.
This contains two buoyancy contributions:  
one related to the density perturbation and a second that we 
named dynamics self gravity (DSG) since it directly links the  
disturbed gravity potential and zonal flows.

The dynamic perturbation of the gravity potential $\potp$ 
is defined by the inhomogeneous differential equation
\bel{eq:HE2}
\left(\nabla^2 + \mu \right) \;\potp = 4\pi\gravc\;\rho^U
\ee
where $\mu$ is the \DSG\ factor and $\rho^U$ is the source term 
describing the impact of the zonal flows. 
The only difference to the classical Poisson equation for  
a gravity potential is the \DSG\ term. 
The dynamic density perturbation $\rho^U$, which is identical 
to the effective density introduced by \citet{Braginsky1995}, 
is obtained from zonal flow and background density 
by a simple integral. 

A polytrope of index unity offers a reasonable approximation 
for the interior of Jupiter and other gas planets. 
This implies that $\mu=\pi^2/R^2$ is constant, which 
considerably eases the task of solving \eqnref{eq:HE2}. 
The problem then assumes the form of an inhomogeneous Helmholtz 
equation and the solution becomes particularly simple when expanding
the radial dependence in modified spherical Bessel functions 
that fulfill the boundary conditions. 
Like in the classical gravity problem, 
Legendre polynomials remain the 
representation of choice for the latitudinal dependence.
These basis functions allow a very efficient (semi) 
analytical solution to the problem. 
Each of the calculations presented here required only a few seconds 
of run time on a standard 4-core notebook. 

There has been a discussion whether
the \DSG\ term could be neglected when inverting high precision gravity 
observations at Jupiter and Saturn for zonal
flow properties. 
Our new formulation allows us to quantify the 
relative impact of the \DSG\ for each gravity harmonic, practically 
independent of the considered zonal flow or background state. 

A special case arises for degree $\ell=1$. 
For the background density with polytropic index unity, 
the $\ell=1$ solution comprises the case where the two buoyancy 
contributions in the TGWE cancel. This  
corresponds to the homogeneous solution of the Helmholtz equation. 
Zonal flow and gravity perturbation then decouple, and it 
becomes impossible to draw on the zonal flows 
from the respective gravity contribution. 
\citet{Kong2017} seem to have noticed the 
related problems without realizing their origin.
However, this is of little practical interest since the origin 
is generally chosen to coincide with the center of gravity
so that $\ell=1$ contributions vanish. 

\Tabref{tab:error} compares the relative \DSG\ impact with 
the precision of newest gravity harmonics of 
Jupiter and Saturn.
The even harmonics $J_2$ to $J_6$ are not listed 
since they are dominated by 
the rotational deformation of the planet.
For Jupiter's $J_3$, $J_5$ and $J_7$ coefficients, 
the relative impact of \DSG\ is comparable to the error and should
thus be taken into account when inverting 
gravity harmonics for zonal flow properties. 
This agrees with the results and conclusion 
by \citet{Kong2017}. 
The error of the higher order 
harmonics may decrease as the 
Juno mission progresses. 
For Saturn, $J_3$, $J_5$ and $J_{10}$ seem precise
enough to warrant including \DSG\ effects. 
The estimates of \citet{Kong2017} and \citet{Galanti2019} 
about the relative impact of the \DSG\ 
is compatible with our results. 
Including the \DSG\ term generally increases the amplitude 
of the gravity coefficients. 

\begin{table}
\centering
\begin{tabular}{c|*{3}{c}}
 $\ell$ & Jupiter & Saturn & $S_\ell$ \\
\hline
 3 &$0.24$ &$0.39$&$0.30$\\
 5 &$0.11$ &$0.24$&$0.13$\\
 7 &$0.14$ &$1.13$&$0.08$\\
 9 &$0.42$ &$0.70$&$0.05$\\
10 &$0.40$ &$0.09$&$0.05$\\
11 &$3.39$ &$1.44$&$0.04$\\
12 &$3.78$ &$0.67$&$0.04$\\
\end{tabular}
\caption{Relative error of gravity harmonics for Jupiter 
\citep{Iess2018} (second column) and Saturn 
\citep{Iess2019} (third column). The fourth column shows 
$S_\ell$, the relative impact of the \DSG\ for 
radial profile $f_3$ also listed in \tabref{tab:ratio}.} 
\label{tab:error}
\end{table}

As pointed out by \citet{Galanti2017} and \citet{Cao2017}, including the rotational 
deformation of the background density in the TWE or TWGE approaches  
may have a similar relative impact on the odd gravity harmonics as the \DSG. 
Both effects may thus have to be taken into account  
when trying to explain these harmonics by the zonal wind dynamics. 

\clearpage
\bibliography{starbound}
\bibliographystyle{elsart-harv}

\renewcommand{\abstractname}{Acknowledgements}
\begin{abstract}
This work was supported by the German Research Foundation (DFG) in the 
framework of the special priority programs 
'Exploring the Diversity of Extrasolar Planets' (SPP 1992). 
\end{abstract}


\appendix

\section{Orthogonality}
\label{sec:ortho}

In this section we show that the spherical Bessel functions
for different $k_{\ell n}$ are orthogonal and that 
$k_{\ell n}^2$ is real.
We start by recalling the properties of a 
self-adjoint or Hemitian linear operator $L$. 
Let $f$ and $g$ be eigenvectors (functions) of $L$ with 
eigenvalues $\lambda$ and $\mu$:
\bel{eq:EV}
L\;f= \lambda\;f \;\;,\;\; L\;g = \mu\;g
\eep
For a self-adjoint operator we have 
\bel{eq:SA}
\langle g , L f\rangle = \langle L g , f \rangle
\eep
It follows that 
\bel{eq:SA2} 
   \lambda\;\langle g , f\rangle = \mu^\star\;\langle g , f\rangle
\ee
and thus $\lambda = \mu^\star$. The eigenvalue is thus 
real and for $\lambda\ne\mu$ we must have
\bel{eq:SA3}
 \langle f , g \rangle =0
\eep
Here the angular brackets denote the integration over the interval
of interest, in our case  
\bel{eq:norm0}
  \langle f ,g \rangle = \int_0^{R}\,d r \;r^2 f^\star\;g
\eep

To show under which conditions an operator is Hermitian, we 
chose a somewhat more general textbook example:
\bel{eq:LT}
   L = a(r) \frac{\partial^2}{\partial r^2}\;+
    \;b(r) \frac{\partial}{\partial r} + c(r)
\eep
Partial integration yields 
\begin{multline}
\label{eq:PI}
\langle f , L\,g \rangle 
     = \left.\left[ r^2 a f \frac{\partial g}{\partial r} 
     +      r^2 b fr g - g \frac{\partial ( r^2 a f )}{\partial r} 
     \right]\right|_{0}^{R} \\ 
     + \int_{0}^{R}\,d r\,g\,
     \left[ \frac{\partial^2 (r^2 a f^\star)}{\partial r^2} - 
            \frac{\partial (r^2  b f^\star)}{\partial r} + r^2 f^\star c \right]
\end{multline}
Rewriting part of the last integral in terms of the 
operator $L$ leads to 
\begin{multline}
\label{eq:PI2}
\langle f , L\,g \rangle = \langle L\,f , g \rangle  \\ + 
\left.\left[ r^2 a f \frac{\partial g}{\partial r} 
     +      r^2 b f g - r^2 a g \frac{\partial f }{\partial r} 
     - f g \frac{\partial ( r^2 a )}{\partial r} 
     \right]\right|_{0}^{r_o} \\ +
     \int_{0}^{r_o}\,d r\ g\,
     \left[ f^\star \frac{\partial^2 (r^2 a)}{\partial r^2} 
     + 2 \frac{\partial (r^2 a)}{\partial r}
     \frac{\partial f^\star}{\partial r}
     - f^\star \frac{\partial (r^2 b)}{\partial r} - 2 r^2 b 
     \frac{\partial f^\star}{\partial r}
     \right]
\end{multline}
The remaining integral vanishes when 
\bel{eq:cond} 
    \frac{\partial (r^2 a) }{\partial r} = r^2 b
\eec
which is certainly the case for the Laplace operator.

The surface contributions only vanish for particular boundary conditions.
When using \eqnref{eq:cond}, the surface contributions vanish for:
\bel{eq:SC}
f \frac{\partial g}{\partial r} 
      - g \frac{\partial f }{\partial r}  = 0 
\eep
There are the three classical options:
\begin{enumerate}
\item Dirichlet boundary conditions $f=0$
\item Neumann boundary conditions $\partial f / \partial r =0$
\item mixed boundary conditions $\partial f/ \partial r + d f =0 $, 
where $d$ is a constant. 
\end{enumerate}
The third option is used for the gravity problem.

We have thus shown that the different eigenfunctions 
defined for each spherical Bessel function $j_\ell(k_{\ell n} r)$ 
(or the second kind $y_\ell(k_{\ell n} r)$) must be orthogonal as long 
as the functions fulfill the boundary conditions. 

\section{Normalization}
\label{sec:norm}

Using 
\bel{eq:diff1}
\langle f, L g \rangle - \langle L f, g \rangle = 
(\mu - \lambda)\,\langle f, g \rangle
\eec
we can define the integral $\langle f, f\rangle$ as the limit
\bel{eq:lim}
 \langle f,f \rangle = \lim_{\lambda\rightarrow\mu} 
 \frac{\langle f,L g\rangle - \langle L f, g \rangle}{\mu-\lambda}
\ee
Using \eqnref{eq:PI2} shows that 
\bel{eq:lim2}
 \langle f,f \rangle = \lim_{\lambda\rightarrow\mu} 
 \frac{\left.\left[ r^2 a f\; \partial g \big/ \partial r 
                \;-\; r^2 a g\;\partial f \big/ \partial r 
      \right]\right|_{r_i}^{r_o}}{\mu-\lambda}
\ee
This limit can be evaluated using l'Hospital's rule.

For the spherical Bessel functions and the Laplace operator 
we are interested in, \eqnref{eq:lim2} reads
\begin{multline}
\label{eq:lim3}
\int_{0}^{r_o}\,d r\;r^2\;j_\ell^2(k r)  = \\
\lim_{k^\prime\rightarrow k} 
 \frac{r_o^2\;\left[j_\ell(k r_o)\; 
 \partial j_\ell(k^\prime r_o) \big/ \partial r  
 \;-\; j_\ell(k^\prime r_o)\;\partial j_\ell(k r_o) \big/ \partial r 
 \right] }{{k}^2-{k^\prime}^2}\;\;,
\end{multline}
where we have used $k=k_{\ell n}$ for brevity.

The result depends on the boundary conditions. 
For the mixed condition the limit becomes 
\begin{multline}
\label{eq:lim8}
\int_{0}^{r_o}\,d r\;r^2\;j_\ell^2(k r) = \\
\lim_{k^\prime\rightarrow k} 
 \frac{r_o^2 j_\ell(k r_o)\;\left[ \partial j_\ell(k^\prime r_o) \big/ \partial r \;+\; (\ell+1)\big/ r_o\;j_\ell(k^\prime r_o)\right]
    }{{k}^2-{k^\prime}^2} \;\;.
\end{multline}
Using recurrence relation \refp{eq:RC1} and L'Hopital's rule yields
\bel{eq:NORM1}
\int_{0}^{R}\,d r\;r^2\;j_\ell^2(k r) = 
 - \frac{ R^3 j_\ell(k R)\;
\partial j_{\ell-1}(k R)\big/ \partial r }{2k^2}
\eep
Finally, using recurrence relations \refp{eq:RC2}  
leads to  
\bel{eq:NORM2}
\int_{0}^{R}\,d r\;r^2\;j_\ell^2(k r) = 
 \frac{R^3}{2}\;j_\ell^2(k R)
\ee
and thus the normalization constant 
\bel{eq:NORM3}
N_{\ell n}= \frac{2^{1/2}}{r_o^{3/2}\,j_\ell(k_{\ell n} r_o)}
\eep

\section{Recurrence relations}
\label{sec:RR}

Some recurrence relations for determining derivatives  
of spherical Bessel functions come in handy.
Standard relations \citep[e.~g.~]{Abramowitz1984} are  
\bel{eq:RC1}
\partial j_\ell(x) \big/ \partial x = j_{\ell-1}(x)\;-\;
(\ell+1)\big/ x\;j_\ell(x)
\eec
and
\bel{eq:RC2}
\partial j_\ell(x) \big/ \partial x = - j_{\ell+1}(x)\;+\; 
\ell\big/ x\;j_\ell(x)
\eep
Combining both allows us to express the second 
derivative as
\bel{eq:RC3}
\partial^2 j_\ell(x) \big/ \partial x^2 = 
-2 \big/ x\;j_{\ell-1}(x)\;-
\left[ 1 - (\ell+1)(\ell+2)\big/ x^2 \right]\;j_\ell(x)
\eep

\section{Equivalence of new and classical solution}
\label{sec:ident}

For $K^2=0$, both the classical solution \eqnref{eq:VE} and 
the new expansion \refp{eq:PE}/\refp{eq:EE2} in spherical Bessel functions should 
be identical.
A comparison shows that this would require 
\begin{multline}
\label{eq:Equi}
\int_0^{r_o}\,d\,\other{r}\;{\other{r}}^{\ell+2}\;{\rhop}(\other{r}) 
\;\overset{?}{=}\;\\
(2\ell+1)\,r_o^{\ell+1}\;\sum_{n=1}^\infty\;\frac{ \cjln(r)}{ \kln^2 }\;
\;\int_0^{r_o}\,d\,\other{r}\;{\other{r}}^2 \cjln(\other{r})\;{\rhop}(\other{r}) \;\;,
\end{multline}
where $\cjln=N_{\ell n} j_\ell(\kln r)$. 

In order to show that this is indeed true, we expand the 
radial dependence under the integral in the classical 
solution into our set of orthonormal spherical Bessel functions:
\bel{eq:EBF}
\other{r}^{\ell} = \sum_{n=1}^\infty\;\cjln(\other{r})\;
\int_0^{R}\,d\,r\;r^{\ell+2}\;\cjln(r)
\eep
Partial integration and using the boundary conditions \refp{eq:BC}  yields
\bel{eq:EBF2}
\int_0^{R}\,d\,r\;r^{\ell+2}\;\cjln(r)= 
\frac{(2 \ell +1)}{\kln}\;\int_0^{R}\,d\,r\;
r^{\ell+1}\;\cj_{\ell-1 n}(r)
\eep
Using recurrence relation \refp{eq:RC1} and performing 
another partial integration finally gives
\bel{eq:Equi2}
\int_0^{R}\,d\,r\;r^{\ell+2}\;\cjln(r) =
 \frac{(2 \ell +1 )}{\kln^2}\;r_o^{\ell+1}\;\cjln(R)
\eep
Plugging this into \eqnref{eq:EBF} and then the result into
the left hand side of \eqnref{eq:Equi} finally proves 
\eqnref{eq:Equi}.


\end{document}